\documentclass[10pt,conference]{IEEEtran}
\IEEEoverridecommandlockouts

\usepackage[utf8]{inputenc}
\usepackage{amsmath, amssymb}
\usepackage{graphicx}
\usepackage[colorlinks=true,urlcolor=teal,
            linkcolor=teal,citecolor=teal]{hyperref}
\usepackage[capitalise]{cleveref}
\usepackage{svg}
\usepackage{xspace}
\usepackage{comment}
\usepackage[expansion,protrusion]{microtype}
\usepackage{tikz}
\usetikzlibrary{arrows.meta, positioning, shapes.geometric, fit, calc}

\newcommand{\eg}[0]{\emph{e.g.\xspace}}

\usepackage[backend=biber, bibencoding=utf8, sorting=none, style=ieee, natbib=true, maxbibnames=3, url=false, doi=false]{biblatex}
\title{Towards System-Level\\ Quantum-Accelerator Integration}

\author{%
\IEEEauthorblockN{Ralf Ramsauer}%
\IEEEauthorblockA{Technical University of Applied Sciences Regensburg\\%
Regensburg, Germany\\%
ralf.ramsauer@oth-regensburg.de}%
\and%
\IEEEauthorblockN{Wolfgang Mauerer}%
\IEEEauthorblockA{Technical University of Applied Sciences Regensburg\\%
Siemens Foundational Technologies\\%
Regensburg/Munich, Germany\\%
wolfgang.mauerer@othr.de}%
}%

\begin{document}
\maketitle
\begin{abstract}
Quantum computers are often treated as experimental add-ons that are  loosely
coupled to classical infrastructure through high-level interpreted languages
and cloud-like orchestration. However, future deployments in both,
high-performance computing (HPC) and embedded environments, will demand tighter
integration for lower latencies, stronger determinism, and architectural
consistency, as well as to implement error correction and other tasks that
require tight quantum-classical interaction as generically as possible.

We propose a vertically integrated quantum systems architecture that treats
quantum accelerators and processing units as peripheral system components. A
central element is the \emph{Quantum Abstraction Layer} (QAL) at operating
system kernel level. It aims at real-time, low-latency, and high-throughput
interaction between quantum and classical resources, as well as robust
low-level quantum operations scheduling and generic resource management. It can
serve as blueprint for orchestration of low-level computational components
``around'' a QPU (and inside a quantum computer), and across different
modalities.

We present first results towards such an integrated architecture, including a
virtual QPU model based on QEMU. The architecture is validated through
functional emulation on three base architectures (x86\_64, ARM64, and RISC-V),
and timing-accurate FPGA-based simulations. This allows for a realistic
evaluation of hybrid system performance and quantum advantage scenarios. Our
work lays the ground for a system-level co-design methodology tailored for the
next generation of quantum-classical computing.

\end{abstract}
\section{Introduction}
Contemporary quantum computers often remain close to physical laboratory
setups, and quantum processing units (QPUs) are connected to classical host
systems through loosely coupled, high-level interfaces. This has enabled rapid
experimental progress, but imposes significant limitations for future
applications, particularly in scenarios that demand scalable, low-latency, and
reproducible quantum-classical interaction.

QPUs will never operate in isolation, but will act as accelerators that are
tightly embedded within heterogeneous classical
systems~\cite{Carbonelli:2024,Yue:2023}. This hybrid nature necessitates a
system architecture that reflects and supports the duality from the ground up.
Future quantum computing deployments will span a wide range of performance
classes: from embedded devices based on compact solid-state technologies, such
as diamond NV centres, to quantum accelerators deployed as extension cards in
workstations, and further to large-scale high-performance quantum computing
(HPQC) clusters in data centres (see \cref{fig:systemgroessen}). These classes
particularly differ in latency constraints and control requirements. A system
architecture must abstract and support all of them without locking into
specific technologies.

Previous efforts, such as QDMI, advocate for more integrated quantum-classical
systems by standardised user-level interfaces. While this is a promising
direction, no such approach does, to the best of our knowledge, take the role
of the operating system kernel into account. We argue that kernel-level
integration is not a mere optimisation, but a necessity.

There are several reasons for this. First, in analogy with classical
accelerators like GPUs, abstraction through kernel-space drivers enables the
decoupling of vendor-specific hardware implementations from standardised
user-space APIs. This is essential for software portability and long-term
ecosystem evolution. Second, kernel-level interaction supports deterministic
scheduling, real-time control, and system-wide resource management. Such
properties are required in embedded control environments and HPC workloads
alike. Third, it allows for hardware abstraction layers that are both
extensible and technology-agnostic, facilitating support for diverse quantum
hardware types, including further emerging quantum technologies such as quantum
sensing or quantum communication.

In contrast to, for example, network-based control interfaces, our design
promotes tight coupling of quantum control electronics to the host system via
standard interconnects. This aims at precise synchronisation, reduced latency,
and unified orchestration under operating system control.

We introduce an integrated QC device model implemented as a virtual accelerator
for early-stage validation. This model, embedded within a full-stack software
environment, dispatches quantum programs from user space through a hardware
abstraction interface. It enables iterative hardware/software co-design and
early functional testing. The architecture is realised using reconfigurable
hardware (FPGA) to simulate time-accurate behaviour and measure key metrics
such as communication latency and scheduling overheads. This allows for
empirical evaluation of quantum systems under near-realistic integration
conditions.
Our main contributions are:

\begin{figure}[htb]
    \includegraphics[width=\linewidth]{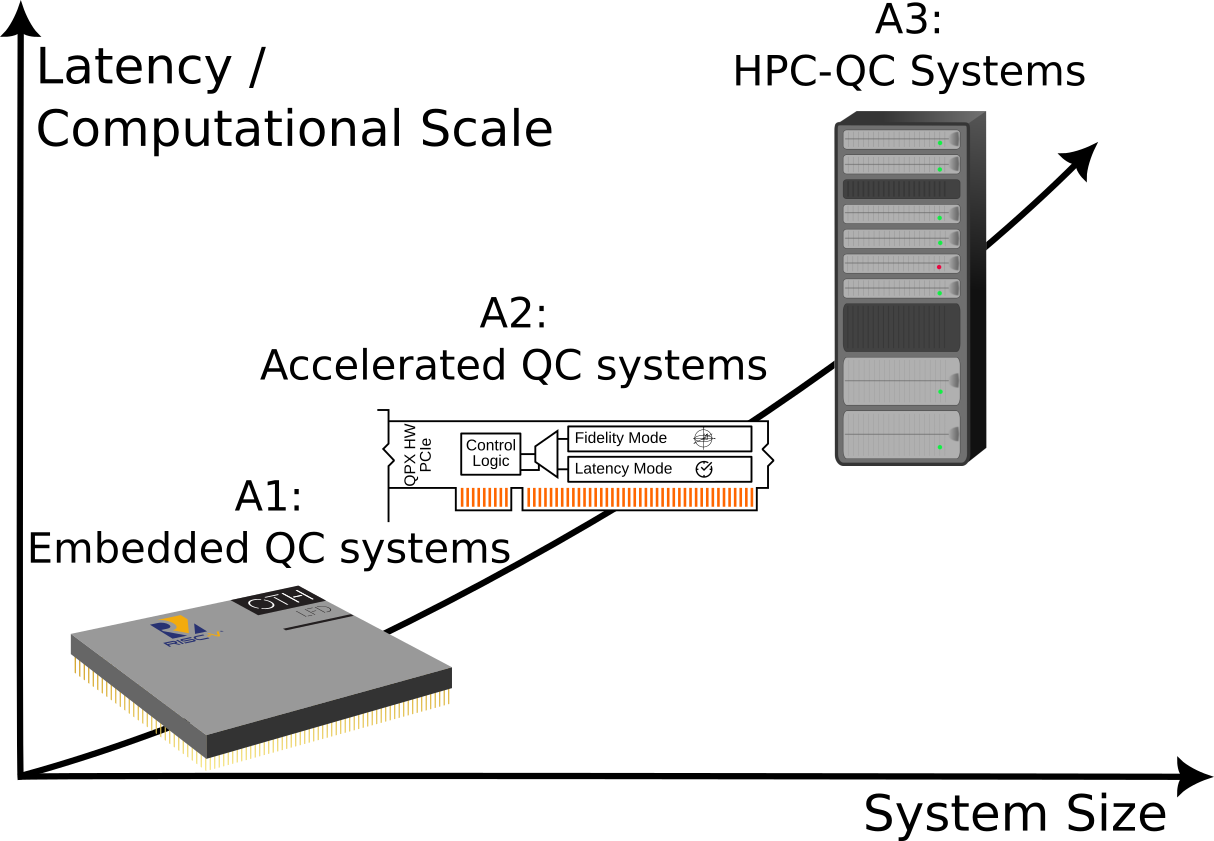}
    \caption{Overview of integration scenarios. We particularly focus on A1 and A2, and consider integration of multiple instances of A2 into A3.}
    \label{fig:systemgroessen}
\end{figure}

\begin{itemize}
\item A tightly coupled quantum-classical system architecture that addresses key requirements for real-time control, portability, and scalability.
\item A working quantum device model using virtualised hardware for early architectural exploration and co-design.
\item We outline a methodology to transition from emulation to time-accurate FPGA-based simulation to analyse real-world integration constraints.
\end{itemize}

\section{State of the Art}

Integration of quantum hardware into classical systems spans multiple
abstraction layers---from high-level algorithm
design~\cite{khan_software_2022,Thelen:2024} to low-level qubit control via
AWGs, RF generators, and FPGAs~\cite{Guo:2024}.  Standardised interfaces and
layered architectures are essential to bridge this gap. Prior work has encoded
low-level quantum instructions as binary
streams~\cite{mccaskey_mlir_2021,stade_towards_2024,fu_eqasm_2019,Guo:2023},
while efforts like QDMI~\cite{wille_qdmi_2024} offer hardware-agnostic
abstractions. Full-stack frameworks have been proposed for hybrid
systems~\cite{Elsharkawy:2025b,zhang2023classical}, yet complete end-to-end
implementations with integrated hardware simulation remain limited.

The RISC-V architecture allows us to make systemic
modifications~\cite{bachrach2012chisel}, and study connection of external
hardware components via FPGA instantiations. While the latter are established
techniques for controlling QPUs for control and
readout~\cite{ryan_2017,xu_2021,Guo:2025}, quantum-classical architectures have
only recently received consideration via quantum instruction
sets~\cite{fu_2017,fu_eqasm_2019,batabyal_2020} or
micro-architectures~\cite{butko_2020,stefanazzi_2022,zhang_2021,Guo:2023,tholen_2022}.
Integrating QPUs into HPC environments has been discussed more
intensively~\cite{Humble_2021,wintersperger:22:codes,safi:23:codesign,Farooqi_2023,seitz_2023},
as summarised in Ref.~\cite{Elsharkawy:2025a}.

\section{Systems Architecture}
\subsection{Overview and Design Rationale}
\paragraph{General Assumptions}
In designing a future-proof architecture for quantum-classical systems, we
start from a set of fundamental assumptions shaped by practical deployment
constraints and technology readiness levels. A significant body of research has
explored quantum extensions to classical instruction set architectures (ISAs),
introducing new instruction formats or opcodes to incorporate quantum-specific
semantics directly into classical compute pipelines. While these approaches are
conceptually appealing and valuable for long-term hardware/software co-design,
we argue that such models are premature for the initial stages of quantum
computing adoption.

In contrast, our architectural perspective is driven by integration pragmatism.
We believe that early quantum computing systems can only be accepted at scale
if they seamlessly integrate into the classical computing environments that are
already in use. In other words, quantum computing components must first become
plug-in extensions to existing compute systems---rather than requiring invasive
or fundamental changes to their core processor architectures. Crucially, these
devices are typically not accessed directly by user-level applications or
libraries. Instead, it is the operating system's task to abstract and
communicate with a QPU controller, a dedicated intermediary that resides either
on the same hardware card or in a tightly coupled component, that provides
means to offload quantum-specific execution, transpilation, translation, and
optimisation tasks. By (optionally) consolidating them in a programmable
controller, the system achieves a high degree of flexibility and autonomy.
This architectural choice reflects the layered nature of the hybrid stack:

General-purpose computation (\eg, application logic, orchestration, scheduling
logic, classical pre-/post-processing) is executed on the host CPU.
Special-purpose operations, such as pulse generation, qubit manipulation, and
physical measurement handling, reside in the quantum control hardware.
Intermediate control logic \emph{may} be offloaded to the QPU controller,
potentially based on flexible and extensible compute architectures such as
RISC-V.

This insight leads to a peripheralisation approach: we assume that
first-generation quantum computing accelerators will be implemented as
classical peripherals---either embedded directly via memory-mapped I/O (MMIO)
interfaces (\eg, in compact setups like diamond-based systems) or as more
general-purpose PCIe-based accelerator cards for workstations and HPC
nodes~(see~\cref{fig:systemgroessen}). Depending on the actual quantum
technology, the actual quantum core may be physically located on the card (as
is conceivable with diamond technologies) or housed externally (\eg, cryogenic
platforms). In the latter case, the interface card acts as a high-speed
low-latency bridge, interfacing with control electronics such as arbitrary
waveform generators (AWGs), microwave electronics, or photonics modules.

\paragraph{Hardware Aspects}
Despite significant physical differences across quantum hardware technologies,
such as superconducting qubits, trapped ions, spin qubits, or NV centers in
diamond, there exist structural commonalities in how these systems are
controlled, measured, and calibrated. These tasks typically involve the
generation of shaped control pulses, precise timing coordination, and
synchronous acquisition of analogue or digital readout signals. Although the
specific waveform characteristics, timing constraints, and physical wiring vary
between modalities, the logical structure of control is strikingly similar
across platforms. This observation enables a generalised control architecture,
which can be tailored through parameterisation rather than restructured from
scratch for each technology.

At a systems level, most quantum control stacks follow a layered pattern
comprising digital signal generation (via CPUs or FPGAs), waveform synthesis
(often via AWGs or DACs), and classical feedback control. By focusing our
design on configurable interfaces and exchange formats, rather than hardwiring
to a specific technology, we enable a hardware abstraction layer that is
general-purpose yet extensible, forming the basis for a reusable and adaptable
stack.

These quantum accelerator cards inherently require special-purpose embedded
control compute to manage configuration, calibration, and runtime orchestration
of quantum operations. The design of such control processors must address
stringent requirements on latency, determinism, and low-level signal fidelity.
To meet these constraints, custom architectural modifications for the
special-purpose control processors are often necessary-for example, to ensure
precise timing alignment or direct hardware-level waveform control. RISC-V
presents a particularly suitable architecture in this context, due to its
openness, modularity, and extensibility. Its structure enables the
implementation of domain-specific extensions tailored to the needs of
quantum-classical interfacing, such as tightly coupled scheduling logic or
application-specific peripheral interfaces. This makes RISC-V an ideal
foundation for the special-purpose embedded control compute required in tightly
integrated quantum systems.

Despite variations in physical deployment, the host interface abstraction
(\eg., MMIO vs.\ PCIe) remains conceptually uniform. This stable abstraction
layer enables portability, system-level validation, and standardisation across
performance classes and quantum technologies.

To rapidly prototype and validate our architectural ideas, we emulate the
accelerator interface using virtualised hardware in QEMU. This model supports
detailed interaction studies, early software stack development, and
architectural experimentation. Communication between host and accelerator
follows state-of-the-art systems techniques, such as MSI-X based interrupt
delivery to ensure scalable, low-latency event signalling from device to host,
DMA-based memory access to enable efficient bidirectional data transfers with
minimal CPU intervention.

Looking forward, our architecture envisions support for multi-QPU
configurations enabling peer-to-peer quantum communication and entanglement
distribution, which are essential for scalable quantum networks and distributed
quantum computing paradigms. To facilitate flexible resource sharing and
isolation, we consider incorporating virtualisation interfaces, such as, for
example Single Root I/O Virtualization (SR-IOV), to provide multiple virtual
quantum accelerator instances on shared physical hardware.

This enables a single physical quantum accelerator device to present multiple
virtual functions, allowing concurrent and isolated access by different host
system components or virtual machines. In high-performance computing (HPC)
environments, this capability facilitates efficient resource sharing and
improved utilisation of costly quantum hardware by multiple users (tenants /
virtual guests) or applications without compromising performance. SR-IOV
reduces overhead by enabling direct device access from user space, bypassing
the hypervisor or kernel layers for critical I/O operations, thereby minimising
latency and maximising throughput. This low-overhead virtualisation is
essential for HPC workloads that require strict timing guarantees and high data
transfer rates, making SR-IOV an attractive approach for integrating quantum
accelerators into large-scale classical-quantum hybrid HPC systems.

Additionally, achieving ultra-low latency and high-throughput interaction
between the classical control units and quantum hardware is critical for
real-time pulse generation and dynamic error correction protocols. Our design
anticipates such tightly coupled co-processing capabilities to enable effective
implementation of these time-sensitive quantum-classical feedback loops.

\paragraph{Software Aspects}

To enable broad applicability and system-level flexibility, our architecture
supports multiple data exchange formats, with a focus on QIR (Quantum
Intermediate Representation) and pulse-level instructions. The rationale for
this dual support lies in the complementary roles of the two representations.
Pulse-level interfaces represent the lowest level of control, offering maximal
flexibility for fine-grained manipulation of quantum operations. This level is
particularly important for experimental setups, calibration procedures, or
future extensions toward quantum-classical co-design at the physical layer.

On the other hand, QIR provides a hardware-agnostic, circuit-level
representation of quantum programs that abstracts from physical details. In our
model, we support direct ingestion of QIR on the accelerator, based on the
assumption that the card itself is best suited to perform transpilation and
optimisation, as it has full knowledge of its internal topology, constraints,
and calibration data. This implies the need for on-card control compute, which
we explicitly respect in our design. The QIR pathway is optional: cards may
accept other formats or perform compilation on the host side if desired.

Future research will explore intelligent caching strategies to minimise
repeated transpilation times and accelerate execution in long-running or
multi-user workloads. The format interface is designed to be extensible,
allowing additional quantum IRs or device-specific formats to be integrated
without architectural changes.

\paragraph{Systems Software Aspects}
To maintain compatibility with emerging and broadly accepted standards, the
software architecture includes a QDMI-style interface embedded into the kernel
driver, allowing integration with higher-level frameworks and runtime systems.

The OS driver-level abstraction must ensure that quantum accelerators can be
managed like conventional hardware resources, enabling multi-process access,
secure context isolation, and system-level orchestration independent of the
scale of the system, from embedded environments to HPC-like environments.
Linux as  basis for implementing our kernel-level driver infrastructure
allows for integration with existing HPC and embedded system software.

The quantum accelerator is exposed as character device (\eg,
\texttt{/dev/qal0}), offering an abstraction that aligns with the standard
device models It is responsible for receiving quantum execution sequences from
user-space applications, queuing and scheduling them according to defined
policies, and tracking execution state. 
Interaction between user space and driver is based on \texttt{ioctl} calls,
providing a flexible mechanism for command invocation and control. Future
revisions will add support for \texttt{mmap}-based interfaces to enable direct
hardware-level access to DMA buffers, allowing low-latency and OS-bypassed
communication paths for applications with determinism requirements. We detail
two instantiations below: a QEMU-based virtual model for rapid iteration and
HW/SW co-design, and a time-accurate FPGA-based hardware-in-the-loop system for
empirical latency and integration analysis.

\subsection{Virtual Quantum Accelerator Model}
To enable rapid prototyping, functional validation, and iterative development,
we have implemented a virtual quantum accelerator using QEMU, a widely adopted
system emulator. This allows us to evaluate architectural concepts and software
interfaces prior to hardware availability, while maintaining tight alignment
with real-world system constraints.

\begin{figure}[htb]
	\begin{center}%
	\fbox{\includegraphics[height=0.14\textheight]{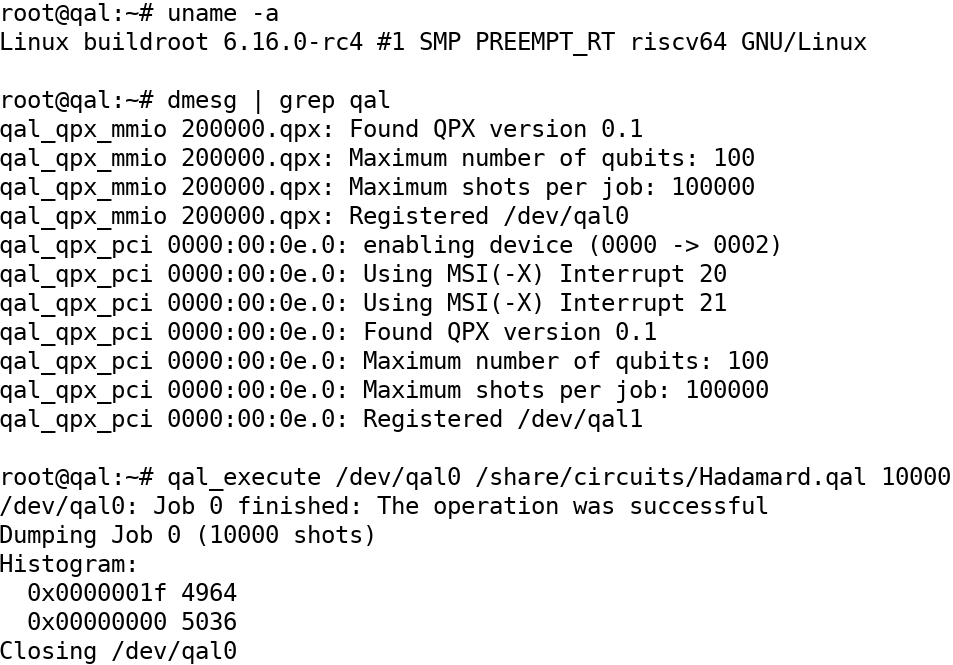}}%
	\fbox{\includegraphics[height=0.14\textheight]{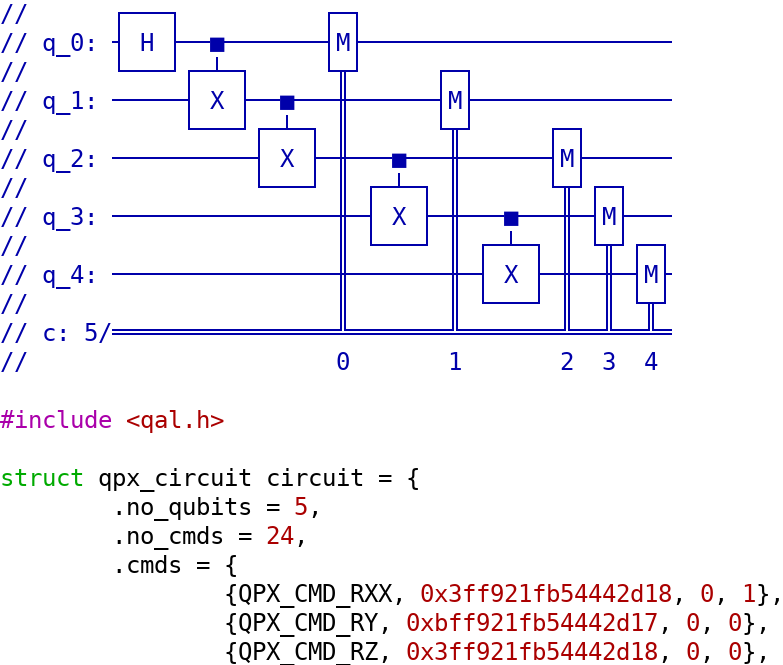}}%
	\end{center}%
	\caption{Illustration of QPX in a virtualised RISC-V environment. Left: Userspace shell within QEMU guest executing a sample quantum circuit via \texttt{/dev/qal0}. Right: Corresponding high-level source (Hadamard.c) generated using Qiskit and compiled into QAL binary format. This demonstrates end-to-end execution through the system stack.
    }
	\label{fig:screenshot}
\end{figure}

Our \emph{Quantum Peripheral Extension} (QPX) acts as drop-in device model for
quantum accelerator modalities. QPX supports MMIO-based embedded integration
and PCIe-based peripheral accelerator configurations, covering a wide spectrum
of possible quantum-classical co-design deployments---from embedded edge
systems to high-performance compute environments. We validate and test our
device model on three different classical host architectures: x86\_64, RISC-V 64
and ARM64.

The device model provides an interface stack, including:
\begin{itemize}
	\item DMA-based memory access for high-throughput, low-latency data exchange,
	\item Support for interrupt signalling via IRQ and MSI(-X),
	\item A command protocol supporting for the internal QAL,
	\item Ongoing integration of QIR (Quantum Intermediate Representation) to facilitate compiler-side integration and future-proofing against emerging quantum software toolchains.
\end{itemize}

The backend uses libquantum as simulation
engine for design validation and ensuring functional correctness from APIs via kernel drivers to device logic. This has several benefits:

\begin{itemize}
	\item Hardware/Software Co-Design: Short development cycles allow architectural feedback from quantum algorithm developers to be directly integrated into interface design.
	\item Early integration testing: Enables system-level validation of driver functionality, kernel interaction, and execution model semantics before hardware availability.
	\item Technology abstraction: Since the model is decoupled from specific quantum hardware implementations, it supports experimentation with technology-agnostic system designs.
	\item Reproducibility~\cite{Mauerer:2022} and CI integration using automated test pipelines.
\end{itemize}

However, QPX cannot be timing accurate, as simulating QPUs efficiently is impossible.
It is not suitable for latency and runtime analysis. 

\subsection{Time-Accurate FPGA Simulation}

To complement the functional validation provided by the QEMU-based virtual
model, we implement a timing-accurate quantum accelerator prototype on a
\emph{digital twin} FPGA platform. This approach enables cycle-accurate and
time-aware simulation of the quantum-classical interface, which is critical for
evaluating latency, throughput, and real-time behaviour. These key factors are
essential for determining potential quantum advantage in hybrid systems.

The FPGA model mirrors the architectural concepts validated in QEMU, supporting
the same PCIe and MMIO interfaces, DMA-based data transfer, and interrupt
handling mechanisms (legacy IRQ, MSI, MSI-X). This ensures a seamless
transition from software simulation to hardware prototyping, facilitating
direct comparison and cross-validation between both platforms.

The ability to operate in two modes is crucial:
\begin{enumerate}
\item Fidelity Mode. The FPGA emulates the quantum hardware (at obvious exponential cost), running quantum circuits as per the virtual device model, verifying correctness of the implementation.
\item Latency Mode. The FPGA mimics timing behaviour of real future quantum hardware components or external devices, providing real-time control and data exchange, enabling in-depth latency profiling and operational testing under realistic conditions.
\end{enumerate}

Reconfigurability of FPGAs enables a flexible, yet precise
platform that balances rapid development cycles with high-fidelity performance metrics. This allows for exploring

\begin{itemize}
	\item quantum-classicalcommunication latencies,
	\item impact of the system / user-level software stack while interacting with real hardware,
	\item real-time constraints for quantum pulse-level control,
	\item overall system behaviour under realistic load conditions.
\end{itemize}

The model bridges between virtual prototypes and physical
hardware, serving as tool for iterative co-design of
hardware and software components, and to fine-tune architectural
parameters for early identification of performance
limitations.

Ultimately, the combination of QEMU virtualisation and FPGA-based
hardware-accurate simulation provides a comprehensive framework for the design,
validation, and optimisation of tightly integrated quantum-classical systems,
driving progress towards practical and scalable quantum accelerators.

\begin{figure}[htbp]
	\vspace*{-0.75em}\includegraphics[width=\linewidth]{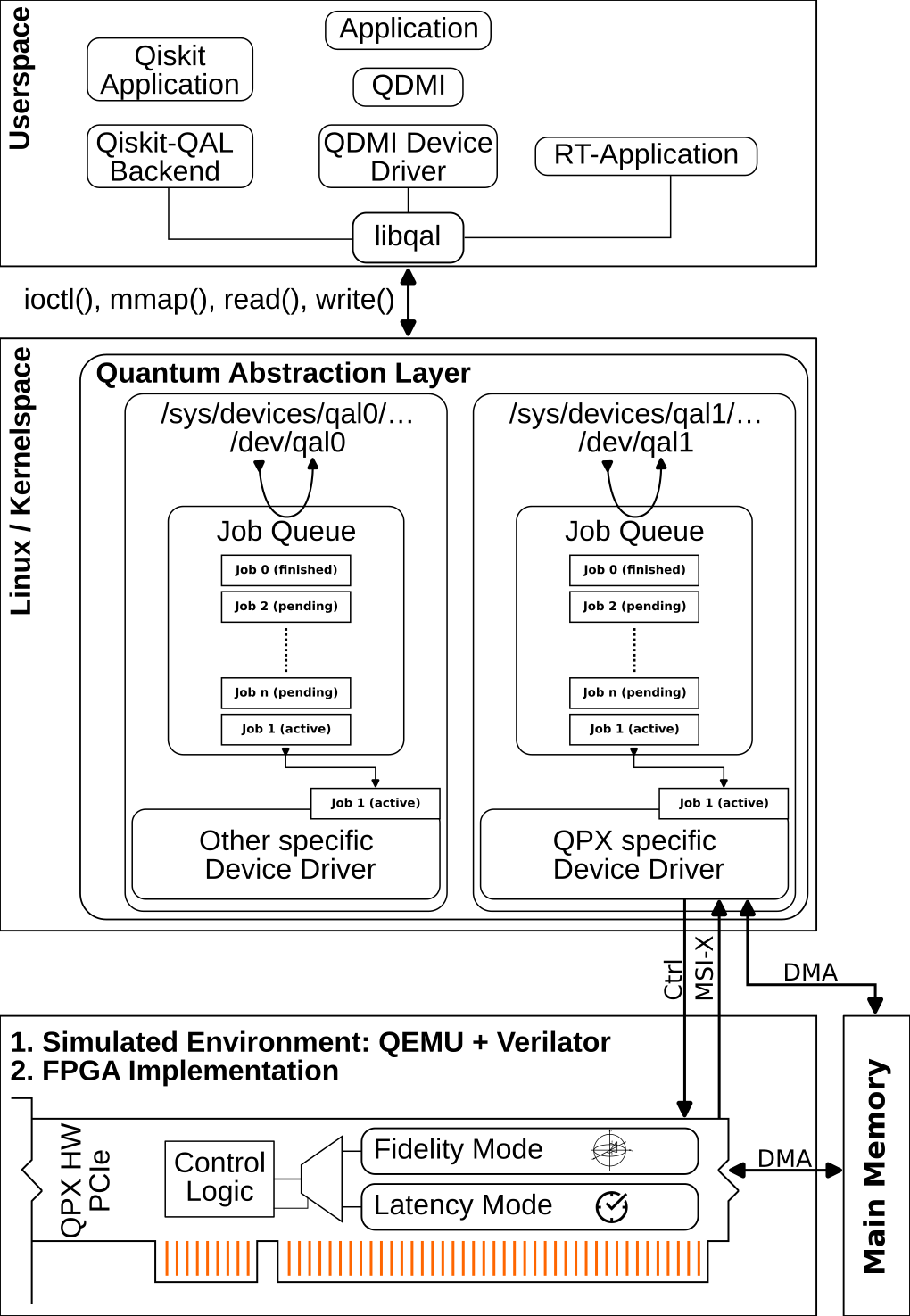}
	\caption{Overview of the prototypical architecture for classical-quantum integration. User space applications interface with standard quantum programming frameworks (\eg, Qiskit, QLM, Qrisp,~\dots), or directly via libqal, a thin user space library providing access to kernel-level quantum device functionality. The Linux kernel hosts a generic QAL subsystem responsible for managing and scheduling quantum sequences, offering abstractions prioritisation and state tracking. This layer connects to device-specific drivers, exemplified by our QPX model, which defines a concrete quantum hardware interface.}
	\label{fig:architecture}
\end{figure}

\section{Prototype Implementation and Roadmap}
The current system is a functionally complete, yet deliberately lean and
modular prototype. All essential components of the classical/quantum
integration stack are realised virtually.

A full cut-through system is demonstrated based on
QEMU (see~\cref{fig:screenshot}), including userspace interaction via the
\texttt{libqal} library, kernel-level job scheduling through a
dedicated device node, and a virtual quantum accelerator model (QPX).
The kernel interface presently uses ioctl-based command handling for task
management and device control. Upcoming versions will include direct
memory-mapped access to DMA regions for low-latency,
trap-free~\cite{ramsauer:17:ospert} interaction.

We validated our stack on three major processor architectures: x86\_64, ARM64,
and RISC-V 64, ensuring portability and architectural agnosticism across
mainstream and emerging compute platforms. These evaluations included both
MMIO-based and PCIe-based configurations, highlighting the versatility of the
QPX model and confirming correct behaviour of the stack across different
execution environments.

While the QEMU-based simulation back-end does not yet support
timing-accurate analysis, it provides an environment for functional testing,
rapid iteration, and architectural prototyping. Insights guide the refinement
of an FPGA-based hardware implementation for cycle-accurate evaluation of
control latencies and communication paths (at the expense of result
correctness).

In summary, the current system validates the feasibility of our integration
model and forms an baseline for subsequent research. It enables
experimentation with interface semantics, abstraction strategies, and
system-level integration patterns in a controlled and especially reproducible
environment. It represents a step towards a broader vision of a
technology-agnostic, tightly integrated classical-quantum computing
architecture.

An important open question that can be empirically addressed concerns division
of responsibilities between kernel and user space. While present interfaces
largely follow a basic pattern (\eg, \texttt{QDMI\_device\_job\_\{submit,
check/wait, get\_results\}}) it remains an open issue which abstractions and
operations should be delegated to kernel-level logic, and which are more
suitably handled in user space.

This becomes increasingly relevant as systems scale. Especially quantum error
correction requires intricate interaction between CPUs, accelerators and QPUs.
Until such mechanisms are fully refined and can be abstracted away, we expect
contradicting requirements between low latency, involved compute, and
flexibility to require system-global architectural decisions. Our architecture
provides the necessary foundation to explore these trade-offs empirically and
systematically. 

\section{Conclusion \& Future Work}

We have introduced a system-level architecture that integrates quantum
accelerators with classical computing environments. Our approach is grounded in
the assumption that hybrid classical-quantum systems use quantum devices
operating as tightly coupled peripherals rather than stand-alone units, and
that the inner working of such stand-alone units features similarities across
modalities that can benefit from a standard base platform.  We have shown that
seamless integration is possible at the kernel level without invasive
modifications.

Our prototypical architecture has been validated across multiple host
architectures (x86\_64, ARM64, and RISC-V 64), and has demonstrated a vertical
system cut-through from user space to simulated hardware. It establishes a
basis for architectural experimentation and refinement, particularly for
challenges like error correction that require the coordinated interaction
between multiple computing entities.

Future work will focus on refining an FPGA-based implementation to enable
time-accurate (and necessarily result-incorrect) evaluation and analysis of
latencies, which is essential for understanding practical quantum advantage
under realistic conditions. We will explore questions related to the placement
of abstractions across user and kernel space, refine the interaction with
emerging interface standards such as QDMI, and investigate advanced offloading
strategies for transpilation and scheduling. Furthermore, virtualisation
capabilities and support for distributed QPU systems will be explored to enable
scalable and shareable quantum-classical infrastructures. The open and modular
nature of our design also supports the release of technology-agnostic
components as open source, thereby contributing to the broader ecosystem and
encouraging future standardisation efforts.

\printbibliography
\end{document}